\documentclass[aps,twocolumn]{revtex4}

\usepackage{epsf}
\usepackage{dcolumn}
\usepackage{longtable}
\usepackage{graphicx}

\begin{document}

\title{Defect free global minima in Thomson's problem of charges on a sphere}

\author{ Eric Lewin Altschuler$^1$, Antonio P\'erez--Garrido$^2$}

\affiliation{$^1$Department of Physical Medicine and Rehabilitation. 
UMDNJ\\
 \hbox{30 Bergen St., ADMC 1, Suite 101 Newark, NJ 07101, USA}\\
email: eric.altschuler@umdnj.edu\\
 $^2$Departamento de F\'\i sica Aplicada, UPCT\\ 
\hbox{Campus Muralla del Mar, Cartagena, 30202 Murcia, Spain}\\
email:Antonio.Perez@upct.es}

\begin{abstract}
Given $N$ unit points charges on the surface of a unit conducting sphere, what configuration of charges minimizes the Coulombic energy $\sum_{i>j=1}^N 1/r_{ij}$? 
Due to an exponential rise in
good local minima, finding global minima for this problem, or even
approaches to do so has proven extremely difficult.  For \hbox{$N = 10(h^2+hk+k^2 )+ 2$}
recent theoretical work based on elasticity theory, and subsequent
numerical work has shown, that for $N \sim >500$--1000 adding
dislocation defects to a symmetric icosadeltahedral lattice lowers the
energy.  Here we show that in fact this approach holds for all $N$, and we
give a complete or near complete catalogue of defect free global minima.
 
\end{abstract}

\maketitle

 What configuration\cite{Th04} of $N$ unit point charges on (the surface of) a unit 
conducting sphere minimizes the Coulombic energy $\sum_{i>j=1}^N 1/r_{ij}$? Beyond physics, 
this questions has utility in understanding the assembly of biological\cite{CK62} 
 and chemical\cite{KH85,LD03} macromolecules, benchmarking 
optimization methods and, in mathematics, Smale\cite{Sm98} has noted the question to be a 
{\it Hilbert} problem for the 21st century.  For $2< N< 100$, the question 
originally posed by Thomson more than a century ago\cite{Th04}, there is agreement  of 
numerical and theoretical work from numerous groups\cite{Wh52,EH91,Ed93,AW94,EH95,HS94,AW97,PO96,MD96,EH97}
using 
a variety of methods so as to have strong confidence that the minimum energy 
configurations have been found.
However, as $N$ grows, due to exponential growth of good local minima\cite{EH95},
 finding global minima has been extremely difficult.  
For $N=10(h^2 + k^2 +hk) + 2$, with $h$ and $h$ integers $h\ge k\ge 0$, highly symmetric icosadeltahedral configurations can be 
constructed (see, e.g., Fig.\ 1).  Initially it was thought that such configurations might be global minima\cite{AW97}, 
but as $N$ grows Dodgson and Moore\cite{DM97} using continuum 
elasticity theory\cite{Do96} suggested that better energy minima could be found 
for $N > \sim 500-1000$ by adding dislocation defects to the icosadeltahedral lattice  (Fig.\ 1).  Indeed, 
this was found to be so\cite{PD97,PD97b,PM99,To,BN00}.  In a full 
census of icosadeltahedral configurations we had recently found that defects lower the 
lattice energy for $N > 792$\cite{AP05}.  We also noted that the theory of Dodgson and 
Moore can also be applied to non-icosadeltahedral defect free configurations.  For example, 
for $N = 78$ a tetrahedral ($T_h$) configuration (Fig.\ 1) is the global energy  minimum\cite{Ed93}, 
and a larger analogue also appears to be the global energy minimum for $N = 306$, see Refs.\ \cite{AW97,AP05}, but for the next larger analogue for $N = 1278$ addition of dislocation defects 
lower the energy\cite{AP05}.  
Here we show that the theory of Dodgson and Moore in fact applies for all $N$, and
give a full or nearly full accounting of defect free configurations for
Thomson's problem.

For each $N$ with a presumed dislocation defect free global minimum\cite{EH91,Ed93,AW94,EH95,HS94,MD96,EH97,Ce} we initially tried 100 trials as such to see if a configuration including dislocation defects with a lower energy could be found: For a given $N$ we started the charges at random locations and minimized the energy  with a standard local gradient descent method.  
If we found a configuration with no dislocation defects and a lower energy than the previously proposed configuration, 
we then tested another 1000 trials to see if a configuration with dislocation defects and lower energy could be found.
One hundred or one thousand trials is hardly even a start to exploring the more than $1.14\cdot 10^6$ predicted\cite{EH95} local minima, for example, for $N =$ 300.  But as we see below, even this few trials yields crucial trends in minima for Thomson's problem.  For some larger $N$, especially those with icosadeltahedral configurations, we  have tried up to 1000 random trials.  Clearly, more extensive trials for all $N$ may give lower energy configurations.

 \begin{figure*}
\begin{center}
\leavevmode
\includegraphics[angle=0,width=12cm]{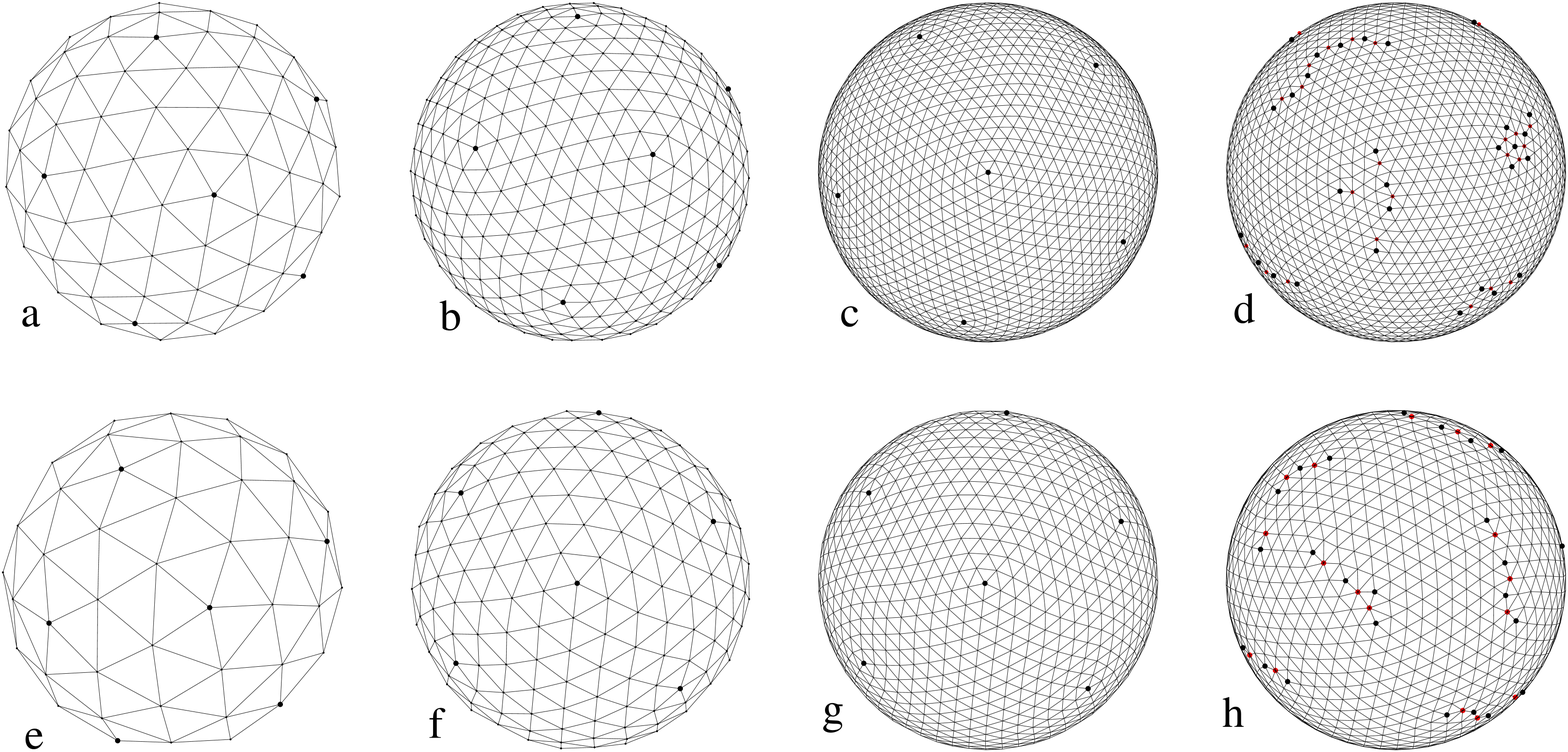}
\end{center}
\caption{
Defect free global energy minimum configurations.
Fivefold coordinated charges (pentamers) are indicated by large black dots, and sevenfold coordinated c
harges (septamers) are indicated by small red dots.  The rest of the charges are sixfold coordinated (hexamers).
a) $N$ = 132 (a (3,1) icosadeltahedral configuration) $E$ = 7875.0453428.
For construction of icosadeltahedral configurations see \cite{AW97}. b) $N =$ 522 (a (6,2) icosadeltahedral
configuration $\equiv 4\times132-6)$, $E$ =129655.8330078.
c) $N$ = 2082 (a (12,4) icosadeltahedral configuration $\equiv=4\times$522-6) $E$ = 2114888.07077971
d)  $N$ = 2082  (defects) $E$ = 2114878.739395074. e) Tetrahedral ($T_h$) configuration for $N = 78$ $E$ =2662.04647456. f)  $T_h$
configuration for $N = 306$ E = 43862.56978081.
g) $T_h$ configuration for $N = 1278$ has a higher energy
718284.03746827, than for a configuration
 h) with dislocation defects 718281.63109628, though we cannot
be certain this is the configuration of minimum energy.
}
\end{figure*}

Our results are summarized in Table I:
For $N =$ 12--200 our search of random configurations confirmed previously found global minima\cite{EH91,Ed93,AW94,EH95,HS94,MD96,EH97,Ce}
(See Ref.\  \cite{Ce} for energies and coordinates, except for $N = 38$ and 46 see \cite{EH95,HS94}.):  For 
$N = 12$--100, there are 81 defect free global minima (91\%). For $N = 101$--200, 92  $N$ have defect free presumed global minima.
For $N = 201$--300 we found 55  $N$ for which the presumed global energy minimum had no dislocation defects.  Of these (see Table II) 12 are $N$ for which the previously presumed global minimum\cite{Ce} also had no dislocation defects but our configuration has a lower energy.
(Contact A.P.G. for coordinates for $N$ listed in Table II; for other $N$ see Ref. \cite{Ce}). There are also three $N$--see Table III-for which a previously presumed global energy minimum had no dislocation defects, but we found a configuration with dislocation defects with a lower energy. 
For example, for $N$ = 214 a defect free configuration had been thought to be the global minimum with energy 21170.0694327506\cite{Ce}, but we found a configuration that has defects with energy 21170.0688491490.
For $N = 301$--400 20  $N$ 
have presumed global energy minimum configurations with no dislocation defects--all previously known 
 (see Ref. \cite{Ce} for energies and coordinates).
For $N = 301$--400 we found one $N$--see Table III-- for which we found a configuration with dislocation defects with lower energy than the previously presumed global energy minimum with no defects.
 For 400 $< N \leq$ 632  we find only eight $N$ for which the presumed global minimum has no dislocation defects and 
 of these eight,  the largest four are icosadeltahedral configurations ($N=$482, 492, 612, 632).  
 We did not find any new configurations 
with no dislocation defects with lower energy than previously presumed global minima, 
but we did find sixteen instances--see Table III--of cases for which the presumed 
global minimum had no dislocation but a configuration with lower energy that includes 
dislocation defects.  These sixteen included, interestingly, two instances--672 and 762--in 
which the presumed global minimum had 
icosadeltahedral symmetry\cite{AP05},
and also $N =$ 542 for which a configuration with high dihedral ($D_5$) symmetry 
had been the presumed global minimum.

\begingroup
\squeezetable
\begin{table}
\caption{$N\le 632$ 
with apparent global minimum energy configurations with no dislocation defects.}
\begin{ruledtabular}
\begin{tabular}{cccc}
12  & 
14--17  & % 
19--20  & %%  7
22--32  \\ %18
34--58 & % 
60--70  &  %52
72--78 &
80--82  \\  % 61
84--108 & % 86
110--122&  % 99
124--125  &  % 101
127--139  \\ %114
141--148  & % 122
150--168  & % 141
170--171  &  %143
173--178  \\  %149
180--200  &  %170
202--210  & % 9
212--213  & % 11
217--226   \\ %21
228--229  &  %23
232  & % 24
234--236  & % 27
239--242  \\ %31
244  & 
246  &
252  & %34
255--258  \\ %38 
260  & 
262  & 
264  & 
266    \\ 
269--270&   
272--273  & %46
276  & 
279  \\ 
282--283 & %50
288--289  & %52
292--293  &
300  \\ %54
302  & %1
304  & 
306 &
312    \\ 
316--317  & %6
322  & 
324  & 
328 \\   %9
348  & 
352  & 
357  & 
361    \\ 
372   &
382 & 
387  & 
390    \\ 
392   &
397  & %19
400 & 
402  \\ 
412    &  
462   &
477   &
482 \\
492 & 
612  &
632   & \\
\end{tabular}
\end{ruledtabular}
\end{table}

 Results for $N = 10(h^2+hk+k^2) + 2$ are summarized in Table IV.
 For $N > 632$ no icosadeltahedral configuration is a global minimum, and
for $N \le 632$ whether or not an icosadeltahedral configuration is a
global minimum depends on the ratio of $h$ to $k$, with smaller ratios protecting
global minima by decreasing energy by rotation of vertices of the
pentamers with respect to each other\cite{AW97,BC02,AP05}.

\begin{table}
\caption{
New apparent global minimum energy configurations  with no dislocation defects.}
\begin{ruledtabular}
\begin{tabular}{ccc}
$N$ & Ref.\ \cite{Ce}  & This work\\
\hline
206 & 19586.024651029       &    19586.023817485   \\
218  &21985.328738558     &      21985.276740701  \\
219  &22191.574733521     &      22191.485474828  \\
229  &24307.641707488        &   24307.607278979 \\
234  & 25401.953728147      &    25401.933332294  \\ 
235  &25623.795960898     &      25623.763144220  \\  
236  &25846.579605445    &       25846.500563170  \\
241  &26975.230903304  &         26975.204068314 \\
246 & 28128.062826837 &       28128.056910358\\
258 & 30994.404751420  &         30994.290832296 \\
264 & 32480.027262398 &          32480.025885504\\
269  &33744.825254911&           33744.824929632  \\
\end{tabular}
\end{ruledtabular}
\end{table}
\endgroup

\begingroup
\squeezetable
\begin{table}
\caption{
Configurations with defects (this work) for
$N$ with previously presumed\cite{Ce} global minima with no dislocation defects. Contact A.P.G.\ for energies and coordinates. Our data will be added to that in Ref.\ \cite{Ce}.}
\begin{ruledtabular}
\begin{tabular}{cccc}
214  &
215  &
227  &
327 \\ 
417  &
447 & 
472& 
512 \\
516 & 
518 &
532 & 
534 \\
537 &
538 & 
542 &   
548 \\
672 &
722 & 
762 & 
777 
\end{tabular}
\end{ruledtabular}
\end{table}
\endgroup
%%%%

 For any $N$ 
for which a defect free configuration appears to be a global minimum we have 
split this $N$ to see if for the next larger analogue the defect free configuration still 
remains an apparent global energy minimum. 
 Configurations are split by putting a new charge midway between each of
the $3N -6$ pairs of charges--for a total of $4N-6$ charges--and then using
a local gradient descent method.
 We continue to split the configuration 
until we found a larger analogue for which the defect free configuration is not 
a global  minimum.  For example, for $N = 78$ it was appreciated some time ago 
that a tetrahedral ($T_h$) configuration was the global energy minimum\cite{Ed93}.
We suggested that for the next larger analogue, $N = 306$, 
the tetrahedral defect free configuration was also a global energy minimum\cite{AW97} 
and this appears to be the case\cite{AP05,Ce}.  But for the next 
larger analogue at $N = 1278$ the tetrahedral configuration  has a higher 
energy than one with dislocation defects.  
Besides $N = 78$ and 306 (78, 306), 
we  have found the following cases in which a split configuration itself also 
appears to be a global energy minimum: (15, 54),  (19, 70),  (25, 94),  (32, 122, 482),  (72, 282) and (77, 302). 

As mentioned above, we had previously thought\cite{AP05} that (137, 542)
was a split pair of likely global minima with high dihedral $D_5$ symmetry.
However, the more trials tested for this paper found that
for $N =$ 542 a configuration with dislocation defects had a lower energy than the
$D_5$, no dislocation defect, analogue of $N = 137$.  Though in the intermediate and somewhat
indeterminate range of the theory of Dodgson and Moore\cite{DM97} --$N \sim$ 500--1000--clearly for
$N$ as small as 542 with a high dihedral, but not icosahedral symmetry, adding dislocation
defects lowers the energy. Also, the global
energy minimum for $N = 522$ is not the icosadeltahedral configuration \cite{Ce}, 
and thus (132, 522) is not a pair of split global minima. However,
 the currently presumed global energy minimum for 522 \cite{Ce}, while
possessing dislocation defects, has twelve defect pairs of a pentamer
and a septamer arranged rather symmetrically and concordantly with the
twelve obligatory pentamers (disclinations).  Thus for $N = 522$, in the
intermediate range for Dodgson and Moore's theory, we see the addition of
defects but in a controlled way.

Table I shows a remarkably strong 
confirmation that the theory
of Dodgson and Moore\cite{DM97} can be applied to general $N$. Not only do dislocation defect free configurations become ever vanishingly rare for $N > 400$, but for $N < 400$--the more so for smaller $N$--the global energy minima typically 
have no dislocation defects. 
Indeed, for $12\leq N\leq 100$ in quite a number of cases special circumstances account for  
presumed global energy minima with dislocation defects: For example, for $N = 13$ it was
 $proven$ many years ago\cite{GM63} that there are no configurations without dislocation defects.  For $N = 18$ the global
minimum configuration has one charge at each pole and four rings of four charges each, staggered with respect to each
other--dihedral $D_{4d}$ symmetry\cite{Ed92}. For $N =$33,  and 79  there seems
to be no way to add one charge, and for $N$ = 71 to subtract one charge,  to the deep global minima for the symmetric configurations of $N = 32$, 72 and 78  and have a good minimum with no dislocation defects.

 Two important questions remain: (1) For the $N$ for which now a defect free configuration is the presumed global minimum (Table I), are these configurations the true global minima?
Given the exponential rise in good local minima with $N$\cite{EH95}, we 
cannot be certain without an amount of numerical testing that exceeds current computational ability, that 
further numerical work may find that some of these configurations are not global minima.
As discussed below, we would expect such instances where defect free configurations fail to be global minima to occur in the 
$\sim 100<N<500$ range.
(2) Are there defect free global energy minimum configurations we have not yet found, either for $N$ not listed in
Table I, or even new lower energy defect free configurations for $N$ in Table I?
Dodgson and Moore\cite{DM97} considered the energy cost of a pair of pentamers in an icosadeltahedral lattice and 
noted that for $N \sim 500$--1000 adding dislocation defects would lower the energy of the overall configuration.  
Numerical work rapidly confirmed this theory\cite{PD97b,PM99,To,BN00,AP05}, and in this work we find that even over 500 
there are at most only two icosadeltahedral configurations that still are possibly global energy minima--though further 
searches on these $N$ may also find these not to be global energy minima.  We noted previously\cite{AP05} that the theory 
of Dodgson and Moore could be applied similarly to a pair of pentamers in a highly symmetric, e.g., tetrahedral, but not 
icosadeltahedral lattice, and similarly (Ref. \cite{AP05} and work above) finds that for $N <$ 500 the symmetric defect 
free configuration appears to be a global energy minimum, but not for $N > 500$.  Here we have pointed out that even for 
general $N$ for a configuration that is dislocation defect free, but not necessarily highly symmetric, still one can use 
the theory of Dodgson and Moore and consider the energy cost of a pair of pentamers.  As the energy cost of a pair of 
vertices will not be lower for a non-symmetric configuration than for a symmetric configuration--as in a non-symmetric 
configuration the cost must be borne of the pair of pentamers with their vertices most closely aligned--the range of 
500--1000 will again be an absolute upper limit of where defect free configurations will remain global minima.  Indeed, 
our numerical work is consistent with the lower range as we have found only six possible defect free configurations 
between 400 and 500.

So by the theory of Dodgson and Moore we don't expect any defect free global energy minima for $N >$ 1000, and likely 
few even in in the $N\sim$500--1000 range. Thus, numerical searches to finalize the catalogue of defect free global 
energy minima should be focused on the $\sim$100--500 range (for  $N \leq 100$ there has been sufficient numerical and 
theoretical work\cite{Wh52,EH91,Ed93,AW94,EH95,HS94,AW97,PO96,MD96,EH97} as to make finding new defect free global energy 
minimum configurations unlikely). In particular, we haven't studied closely yet those $N$ for which the currently 
proposed\cite{Ce} global minimum includes dislocation defects. For these $N$ more numerical trials could find better minima 
that have no dislocation defects.

\begingroup
\squeezetable
\begin{table}
\caption{
Energy of icosadeltahedral configurations.  An * 
indicates a non-icosadeltahedral configuration (with or without defects) of lower energy, though not necessarily the global minimum. }
\begin{ruledtabular}
\begin{tabular}{ccd}
$N$ &  $h,k$  & \mbox{Energy} \\
\hline
          12
 &
 h=           1 k=           0
 &   49.165253058     \\
          32
 &
 h=           1 k=           1
 &   412.261274651     \\
          42
 &
 h=           2 k=           0
 &   732.256241038     \\
 &
*Non--icosadeltahedral
 &   732.078107551     \\
          72
 &
 h=           2 k=           1
 &   2255.00119099     \\
          92
 &
 h=           3 k=           0
 &   3745.618739085     \\
 &
*Non--icosadeltahedral
 &   3745.291636245     \\
         122
 &
 h=           2 k=           2
 &   6698.374499261     \\
         132
 &
 h=           3 k=           1
 &   7875.045342816     \\
         162
 &
 h=           4 k=           0
 &   11984.551433873     \\
 &
*Non--icosadeltahedral
 &   11984.050335831     \\
         192
 &
 h=           3 k=           2
 &   16963.338386471     \\
         212
 &
 h=           4 k=           1
 &   20768.053085969     \\
         252
 &
 h=           5 k=           0
 &         29544.282192861     \\
&
*Non--icosadeltahedral
&         29543.528647529          \\
         272
 &
 h=           3 k=           3
 &   34515.193292688     \\
         282
 &
 h=           4 k=           2
 &   37147.294418474     \\
         312
 &
 h=           5 k=           1
 &   45629.362723819     \\
         362
 &
 h=           6 k=           0
 &   61720.023397813     \\
 &
* w/defects
 &   61719.309054516\footnotemark[1]  \\
         372
 &
 h=           4 k=           3
 &   65230.027122566     \\

         392
 &
 h=           5 k=           2
 &   72546.258370895     \\
         432
 &
 h=           6 k=           1
 &   88354.229380725     \\
 &
* w/defects
 &   88354.190665226\footnotemark[1]    \\
         482
 &
 h=           4 k=           4
 &   110318.139920155     \\
         492
 &
 h=           7 k=           0
 &   115006.982258289     \\
 &
h=           5 k=           3
 &   115005.255889700     \\
         522
 &
 h=           6 k=           2
 &   129655.833007858     \\
 &
* w/defects
 &   129655.326253464\footnotemark[2]   \\
 
         572
 &
 h=           7 k=           1
 &   156037.879346228     \\
 &
* w/defects
 &   156037.222417655\footnotemark[2]   \\
         612
 &
 h=           5 k=           4
 &   178910.494981768     \\
         632
 &
 h=           6 k=           3
 &    190937.233325601    \\
         642
 &
 h=           8 k=           0
 &   197100.363816212     \\
 &
* w/defects
 &   197098.532524683\footnotemark[2]   \\
         672
 &
 h=           7 k=           2
 &   216171.432658341     \\
 &
 * w/defects
 &216171.227524558\footnotemark[3]\\
 
         732
 &
 h=           8 k=           1
 &   256975.527362500     \\
 &
* w/defects
 &  256973.838562012 \footnotemark[2]    \\
         752
 &
 h=           5 k=           5
 &   271362.588212841     \\
 
 &
 * w/defects
 &271361.125880198\footnotemark[2] \\
         762
 &
 h=           6 k=           4
 &   278704.548699996     \\
  &  
  * w/defects
  & 278704.428077126\footnotemark[3]
  \\ 
         792
 &
 h=           7 k=           3
 &   301321.818305597     \\
 
 &
 * w/defects
 &301320.370436992\footnotemark[2] \\

\end{tabular}
\end{ruledtabular}
 \footnotetext[1]{Ref.\ \cite{AP05}}
  \footnotetext[2] {Ref.\ \cite{Ce}}
  \footnotetext[3] {This work}

\end{table}
\endgroup

\begin{figure}
\begin{center}
\leavevmode
\includegraphics[angle=-90,width=9.5cm]{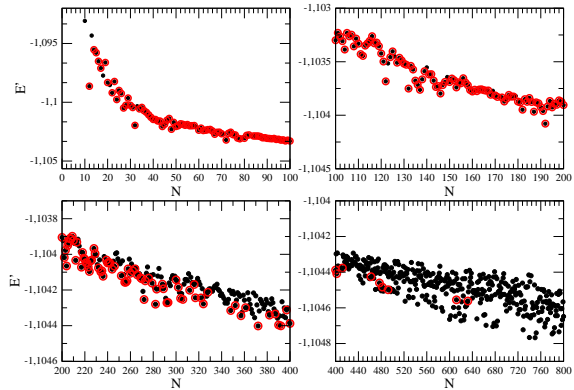}
\end{center}
\caption{
$E^\prime=(2E-N^2)/N^{3/2}$  {\it vs.\/} $N$. 
$E^\prime$ is defined using Eq.\ (\ref{ecuacion}). 
Energies are plotted with black dots 
and are encircled in red if the associated presumed global minimum energy configuration has no dislocation defects.}
\end{figure}

For another reason we think that if new defect free global energy minimum energy configurations are to be found one must look for $N <$ 500. Using the method of Ewald sums\cite{PM99,To,BM77}, one finds that the energy of $N$ charges on a unit sphere in the theoretically impossible (by Euler's theorem), but approximately useful, construct of a perfect triangular (hexagonal) lattice for 
$N\rightarrow \infty$  is
\begin{equation}
E=\frac{1}{2}\left( N^2-1.106 103 3 N^{3/2}\right)
\label{ecuacion}
\end{equation}
where the term order $N^{3/2}$ is the energy of $N$ charges uniformly distributed on a sphere and embedded in a uniform neutralizing background\cite{BM77} and the term  $N^2/2$ accounts for the lack of a uniform neutralizing background in Thomson's problem. Eq.\ (\ref{ecuacion}) has been also obtained using other techniques by a number of  authors\cite{Wa90,Wa92,RS94}.
As $N$ grows large, in accordance with equation (1), $E^\prime=(2E-N^2)/N^{3/2}$ approaches $-1.11061033$.
Previous numerical calculations for $N\leq 200$ yielded a value  -1.1046 for the constant coefficient of the $N^{3/2}$ term\cite{RS94,GE92}, though this is clearly seen to be exceeded for $N>\sim 600$ (Fig.\ 2). Furthermore,
a configuration with $N= 15152$ and $E^\prime=-1.10562321$ has been found\cite{PM99}.  
$E'$ is plotted in Fig.\ 2.  We see that for $N < 500$ the defect free energy configurations stand out as having particularly low relative scaled energies, while for $N > 500$ the defect free configurations are not particularly good compared with other presumed global energy minimum configurations. Thus, for $N > 500$ even for the currently presumed defect free global energy minimum configurations there seems no added benefit compared to configurations with defects, and thus we doubt that for other $N$ in this range defect free configurations will be global minima.

%A.P.G. would like to acknowledge  financial support from Spanish MCyT
%under grant No. MAT2003--04887.

\end{document}